\documentclass[aoas,nameyear,seceqn,dvips]{arximspdf}
\usepackage{dcolumn}
\usepackage{graphicx}

\doi{10.1214/10-AOAS361}
\volume{4}
\issue{2}
\pubyear{2010}
\firstpage{715}
\lastpage{742}

\makeatletter
\DeclareMathAlphabet\mathcaligr{OMS}{cmsy}{m}{n}
\renewcommand{\mathcal}{\mathcaligr}
\newcommand{\eqref}[1]{(\ref{#1})}
\renewcommand{\cite}{\citet}
\newcommand{\Esp}{\mathbb{E}}
\newcommand{\Hcal}{\mathcal{H}}
\newcommand{\Jcal}{\mathcal{J}}
\newcommand{\Mcal}{\mathcal{M}}
\newcommand{\Pcal}{\mathcal{P}}
\newcommand{\RX}{R_{\Xbf}}
\newcommand{\wRX}{\widehat{R}_{\Xbf}}
\newcommand{\alphabf}{\bolds\alpha}
\newcommand{\Psibf}{\bolds\Psi}
\newcommand{\gammabf}{\bolds\gamma}
\newcommand{\thetabf}{\bolds\theta}
\newcommand{\taubf}{\bolds\tau}
\newcommand{\Zbf}{\mathbf{Z}}
\newcommand{\Xbf}{\mathbf{X}}
\newcommand{\ybf}{\mathbf{y}}
\newcommand{\Zerobf}{\mathbf{0}}
\newcommand{\Pro}{\mathbb{P}}
\newcommand{\Bcal}{\mathcal{B}}
\newcommand{\pbf}{\mathbf{p}}
\newcommand{\Ncal}{\mathcal{N}}
\newcommand{\mubf}{\bolds\mu}
\newcommand{\Sigmabf}{\bolds\Sigma}
\newcommand{\Wbf}{\mathbf{W}}
\newcommand{\Ibb}{\mathbb{I}}
\newcommand{\betabf}{\bolds\beta}
\newcommand{\Ybf}{\mathbf{Y}}
\newcolumntype{d}[1]{D{.}{.}{#1}}
\makeatother

\begin{document}
\begin{frontmatter}

\title{Uncovering latent structure in valued graphs: A~variational approach}
\runtitle{Latent structure in valued graphs}

\begin{aug}
\author[A]{\fnms{Mahendra} \snm{Mariadassou}\ead[label=e1]{mariadas@agroparistech.fr}},
\author[A]{\fnms{St\'ephane} \snm{Robin}\corref{}\ead[label=e2]{robin@agroparistech.fr}}
\and
\author[C]{\fnms{Corinne} \snm{Vacher}\ead[label=e3]{corinne.vacher@pierroton.inra.fr}}
\runauthor{M.~Mariadassou, S. Robin and C. Vacher}
\affiliation{AgroParisTech and INRA, AgroParisTech and INRA\\ and
INRA and University Bordeaux I}
\address[A]{
M. Mariadassou\\
S. Robin\\
UMR 518 AgroParisTech/INRA MIA\\
16, rue C. Bernard, F-75005 Paris\\
France\\
\printead{e1}\\
\phantom{E-mail: }\printead*{e2}} 
\address[C]{C. Vacher\\
UMR 1202 Univ. Bordeaux I/INRA BioGeCo\\
69, route d'Arcachon, F-33612
Cestas\\
France \\
\printead{e3}}
\end{aug}

\received{\smonth{10} \syear{2008}}
\revised{\smonth{4} \syear{2010}}

%
\begin{abstract}
As more and more network-structured data sets are available, the
statistical analysis of valued graphs has become common place.
Looking for a latent structure is one of the many strategies used
to better understand the behavior of a network. Several methods
already exist for the binary case.

We present a model-based strategy to uncover groups of nodes in
valued graphs. This framework can be used for a wide span of
parametric random graphs models and allows to include covariates.
Variational tools allow us to achieve approximate maximum
likelihood estimation of the parameters of these models. We
provide a simulation study showing that our estimation method
performs well over a broad range of situations. We apply this
method to analyze host--parasite interaction networks in forest
ecosystems.
\end{abstract}

%

\end{frontmatter}

\section{Introduction}

Data sets presenting a network structure are increasingly studied in many
different domains such as sociology, energy, communication, ecology or
biology [\cite{AlB02}]. Statistical tools are therefore needed to analyze
the structure of these networks, in order to understand their
properties or
behavior. A strong attention has been paid to the study of various
topological characteristics such as degree distribution, clustering
coefficient and diameter [see, e.g., \cite{BaA99}, \cite{NWS02}]. These
characteristics are useful to describe networks but not sufficient to
understand its whole structure.

A natural and intuitive way to capture an underlying structure is to look
for groups of edges having similar connection profiles [\cite{GeD04},
\cite{NWS02}], which is refereed to as community detection
[\cite{GiN02}, \cite{New04}]. This usually turns into an unsupervised
classification (or clustering) problem which requires efficient estimation
algorithms since the data set at hand is ever increasing.

Several attempts at community detection have been proposed in the
literature: greedy algorithms for community detection [\cite{GiN02} and
\cite{New04}] and clustering based on spectral analysis of the adjacency
matrix of a graph [\cite{LBB07}]. Greedy algorithms and spectral
clustering both assume that communities are determined by a strong
within connectivity opposed to a low between connectivity. This might
be true for so-called communities but need not be true for other
groups of nodes. For example, a group of nodes loosely connected to
each other but highly connected to a specific group of hubs have the
same connection profile and form a homogeneous group but do not form
a community. In addition, they do not offer an explicit generative
model nor a criterion to select the correct number of communities.

Model-based methods are appealing by contrast: explicit modeling of the
heterogeneity between nodes gives different groups an intuitive and
easy to
understand interpretation. Several probabilistic models exists for random
graphs [see \cite{PaR07} for a complete review], ranging from the seminal
Erd\"os--R\'enyi (ER) model [\cite{ErR59}] to the sophisticated Stochastic
Block Model (SBM) [\cite{NoS01}]. The ER model assumes independent and
identically distributed edges which entails that all nodes are structurally
equivalent and, thus, there is only one community, although a big one. The
$p_1$ model from \cite{HoL81} extended the ER model by assuming independent
\textit{dyads} instead of \textit{edges}, allowing the breakthrough from
undirected to directed graphs. But again, all nodes are structurally
equivalent in the $p_1$ model. \cite{FiW81} and \cite{FMW85} lifted these
constraints by assuming the nodes are distributed among $Q$ classes with
different connectivity profiles. In this model, groups are easily
interpreted as nodes belonging to the same class. Unfortunately,
\cite{FMW85} assumes class assignments are perfectly well known, which
rarely happens. The state of the art in terms of graph modeling is the SBM,
inspired by \cite{LoW71} and introduced by \cite{NoS01}, which takes
advantage of mixture models and unknown latent variables to allow an easy
modeling of groups without requiring them to be known in advance.

In the SBM framework, community detection boils down to three crucial
steps: assignment of nodes to groups, estimation of the model parameter and
selection of the correct number of groups. Several authors offered their
method to solve these issues using Bayesian methods. \cite{NoS01} work with
the original SBM model. \cite{HoW08} work in a highly constrained version
of SBM in which heterogeneity is strictly limited to intra- and
inter-community connection and thus characterized by only two parameters,
against $Q^2$ in the unconstrained SBM. \cite{ABF08} extend the SBM
framework by allowing nodes to exhibit multiple communities. By contrast,
\cite{DPR08} use a frequentist approach to estimate the parameters of the
SBM. The frequentist approach is less computation intensive than its
Bayesian counterpart, whereas the Bayesian approach is supposed to better
account for the uncertainty. With the notable exception of \cite{NoS01},
who use MCMC to estimate the model parameter, both lines of work make heavy
use of variational techniques: either Variational EM [\cite{Jaa00}] or
Variational Bayes [\cite{Att00}; \cite{BeG03}; \cite{XJR03}; \cite{WBJ05}]. MCMC computational cost
is prohibitive, effectively leading to severe size limitations (around 200
nodes). Furthermore, because of the complex likelihood landscape in the
SBM, good mixing of the Markov Chain is hard to achieve and
monitor. Variational approximations, by contrast, replace the
likelihood by
a simple surrogate, chosen so that the error is minimal in some
sense. Frequentist and Bayesian approach then differ only in the use of
this surrogate likelihood: Bayesians combine it to a prior distribution of
the parameter (chosen from some suitable distribution), whereas frequentists
use it directly. In all these methods, the number of groups is fixed during
the estimation procedure and must be selected using some criterion. By
contrast, \cite{KGT04} propose an original approach where the number of
groups changes and is selected \textit{during} the estimation process. Both
Bayesian and frequentist estimations approaches give the same kind of
results: an optimal number of groups and a probabilistic assignment of
nodes to groups, depending on their connection profile. However, the
Bayesian estimation strategy leads to severe constraints on the choice of
prior and hyperprior distributions. The \cite{DPR08} maximum likelihood
approach does not require any prior specification and is more efficient
than MCMC estimation [\cite{PDM07}].\looseness=1


Previous models are all models for binary networks, for which the only
information is the presence or absence of an edge. Binary information
certainly describes the topology of a network but is a rather poor
description. It accounts neither for the intensity of the
interaction between two nodes nor for the specific features of an
edge. The intensity of an edge may typically indicate the amount of
energy transported from one node to another, the number of passengers
or the number of common features between two nodes, whereas the
specific feature of an edge may be the phylogenetic distance between
its two ending nodes. Many networks, such as power, communication,
social, ecological or biological networks, are naturally valued and
are somehow arbitrarily transformed to a binary graph. This
transformation sometimes conceals important results [\cite{TTL07}].
Extending binary models and the associated estimation procedures to
valued graphs with specific features allows more complexity, and more
relevant information with it, to be processed while estimating the
structure of the network.

We are motivated by the search of a structure in valued graphs
describing the similarity between species within an assemblage
according to their biotic interactions. In ecology, an assemblage is
defined as a taxonomically related group of species that occurs in the
same geographic area [\cite{Rim00}]. The species composing an
assemblage usually interact with many species belonging to other
assemblages and the nature of these interactions is often very diverse
(predator--prey interactions, host--parasite interactions, mutualistic
interactions, competitive interactions). One of the questions facing
ecologists is to understand what determines with whom a species
interact. Conventional wisdom is that within an assemblage, two
closely related species should share more interactions than two
evolutionary distant species because the range of interactions of a
species is constrained by its physiological, morphological and
behavioral attributes. In several cases, this conventional wisdom
is revealed to be true. Phylogenetically related plant species have been
shown to bear similar pathogens and herbivores [\cite{BrB06}; \cite{GiW07}]
and the diet's range of predators has been shown to be
phylogenetically constrained [\cite{CBB04}]. This tendency for
phylogenetically related species to resemble each other is called
phylogenetic signal [\cite{BlG02}]. In other cases, no phylogenetic
signal was detected [\cite{RLG07}; \cite{VPD08}]. Selection pressures exerted
by the environment might account for this absence: species have to
adapt to varying environments to survive, diverging from close
relatives in their physiology, morphology and behavior, and possibly
developing novel interactions [\cite{BeK08}; \cite{CBB04}]. The valued graphs
under study have species as nodes and the number of shared
interactions as edges. We use a mixture model with phylogenetic
distance between species as covariate to measure the strength of the
phylogenetic signal. This latter is defined as the decrease in the
number of selected groups due to the inclusion of the covariate. Two
different assemblages are considered. The first assemblage is composed
of 51 tree species occurring in the French forests and the second is
composed of 153 parasitic fungal species also occurring in the French
forests. The interactions considered are host--parasite interactions.
We expect to find a lower phylogenetic signal in the host range of
parasitic fungal species [\cite{BeK08}; \cite{RMA06}; \cite{VPD08}]
than in the
vulnerability of tree species to parasites [\cite{BrB06}; \cite{GiW07};
\cite{VPD08}].


In this paper we propose an extension to the stochastic block model,
introduced in \cite{FiW81}; \cite{FMW85}; \cite{NoS01}, and the methods
of \cite{AiC05}
and \cite{DPR08}, that deals with valued graphs and accounts for
possible covariates. We use a general mixture model describing the
connection intensities between nodes spread among a certain number of
classes (Section \ref{Sec:Model}). A~variational EM approach to get an
optimal, in a sense to be defined, approximation of the likelihood is
then presented in Section \ref{Sec:LikeVar}. In Section
\ref{Sec:Estim} we give a general estimation algorithm and derive some
explicit formulas for the most popular distributions. The quality of
the estimates is studied on synthetic data in Section \ref{Sec:Simul}.
Finally, the model is used to elucidate the structure of host--parasite
interactions in forest ecosystems and results are discussed in Section
\ref{Sec:Ecology}.

\section{Mixture model} \label{Sec:Model}

We now present the general extension of SBM to valued graphs and
discuss the two particular modelings used for the tree species and
fungal species interaction networks.

\subsection{Model and notation}

\begin{description}
\item[Nodes.] Consider a graph with $n$ nodes, labeled in
$\{1,\ldots,n\}$. In our model the nodes are distributed among
$Q$ groups so that each node $i$ is associated to a random vector
$\Zbf_i = (Z_{i1},\ldots,Z_{iQ})$, with $Z_{iq}$ being $1$ if node
$i$ belongs to group $q$ and $0$ otherwise. The $\{\Zbf_i\}$ are
supposed to be independent identically distributed observations from
a multinomial distribution:
%
%
\begin{equation} \label{Eq:DistZ}
\{\Zbf_i\}_i \mbox{ i.i.d.} \sim\Mcal(1; \alphabf),
\end{equation}
where $\alphabf= (\alpha_1,\ldots,\alpha_Q)$ and $\sum_q \alpha_q
= 1$.
\item[Edges.] Each edge from a node $i$ to a node $j$ is
associated to a random variable $X_{ij}$, coding for the strength of
the edge. Conditionally to the group of each node, or
equivalently knowing the $\{\Zbf_i\}$, the edges are supposed to be
independent. Knowing group $q$ of node $i$ and group $\ell$ of
node $j$, $X_{ij}$ is distributed as $f(\cdot,\theta_{q\ell}) :=
f_{q\ell}(\cdot)$, where $f_{\theta_{q\ell}}$ is a probability
distribution known up to a finite-dimensional parameter
$\theta_{q\ell}$:
%
%
\begin{equation} \label{Eq:DistX}
X_{ij}|i \in q,j \in\ell\sim f(\cdot,\theta_{q\ell}) := f_{q\ell
}(\cdot).
\end{equation}
\end{description}
Up to a relabeling of the classes, the model is identifiable and
completely specified by both the mixture proportions $\alphabf$ and
the connectivity matrix $\thetabf= (\theta_{q\ell})_{q,\ell= 1,
\ldots,
Q}$. We denote $\gammabf= (\alphabf,\thetabf)$ the parameter of the
model.

\subsubsection*{Directed and undirected graphs}
This modeling can be applied to both directed and undirected graphs.
In the directed version, the variables $X_{ij}$ and $X_{ji}$ are
supposed to be independent conditionally to the groups to which
nodes $i$ and $j$ belong. This hypothesis is not always realistic
since, for example, the traffic from $i$ to $j$ is likely to be
correlated to the traffic from $j$ to $i$. A way to account for such
a dependency is to consider a undirected graph with edges labeled
with the bivariate variables $\{(X_{ij}, X_{ji})\}_{1 \leq i < j
\leq n}$. All the results presented in this paper are valid for
directed graphs. The results for undirected graphs can easily be
derived and are only briefly mentioned.



\subsection{Modeling the number of shared hosts/parasites}
\label{subsec:PoisModel}

In our tree interaction network, each edge is valued with the number
of common fungal species two tree species can host. Our purpose is to
understand the structure of this network and it is natural to model
the counts $X_{ij}$ as Poisson distributed. The mixture models aims
at explaining the heterogeneity of the $X_{ij}$. However, we would
also like to account for some factors that are known to be
influential. In our network, we expect two phylogenetically related
tree species $i$ and $j$ to share a high number $X_{ij}$ of parasitic
species. As such, their average number of shared parasitic species
$\Esp[X_{ij}]$ is expected to decrease with their phylogenetic
distance $y_{ij}$. We consider three alternatives, and compare two of
them.
%
\begin{description}
\item[Poisson mixture (PM):] In this mixture, we do not account for the
covariates and $X_{ij}$ only depends on the classes of $i$ and $j$:
\[
X_{ij}|i \in q, j \in\ell\sim\Pcal(\lambda_{q\ell}).
\]
$\lambda_{q\ell}$ is then the mean number of common fungal species
(or mean interaction) between a tree species from group $q$ and one
from group $\ell$ and $\theta_{q\ell} = \lambda_{q\ell}$.
\item[Poisson regression mixture with inhomogeneous effects (PRMI):] In
this mixture, we account for the covariates via a regression model that
is \textit{specific} to the classes of $i$ and $j$:
\[
X_{ij}|i \in q, j \in\ell\sim
\Pcal(\lambda_{q\ell}e^{\beta_{q\ell}^{\intercal}\ybf_{ij}}),
\]
where $\ybf_{ij}$ is a vector of covariates and $\theta_{q\ell} =
(\lambda_{q\ell}, \beta_{q\ell})$.
\item[Poisson regression mixture with homogeneous effects (PRMH):] In
this mixture, the effect of the covariates does not depend on the classes
of $i$ and $j$:
\[
X_{ij}|i \in q, j \in\ell\sim
\Pcal(\lambda_{q\ell}e^{\beta^{\intercal}\ybf_{ij}}),
\]
$\theta_{q\ell} = (\lambda_{q\ell}, \beta)$.
\end{description}

We point out that models PRMI and PRMH have different purposes. In PRMI,
the link between the covariates and the edges is locally refined
\textit{within} each class $(q,\ell),$ whereas in PRMH, the
covariates compete
globally with the group structure found by PM. In PRMH, the mixture looks
for remaining structure among the residuals of the regression model. If the
structure was completely explained by the covariates, the possibly many
components found using PM would reduce to a single component when using
PRMH. To a lesser extent, we expect the number of components to be smaller
with PRMH than with PM if the phylogenetic distance explains part of the
structure. As we look for structure beyond the one already explained by the
covariates, we consider only models PM and PRMH.

The same models are used for the fungal species interaction network.
In our examples, data consist in counts, but other types of data can be
handled with similar mixture and/or regression models (see Appendix
\ref{SubSec:ClassicDist} for details).

\section{Likelihood and variational EM} \label{Sec:LikeVar}

We now address the estimation of the parameter $\gammabf=
(\alphabf, \thetabf)$. We show that the standard maximum likelihood
approach cannot be applied to our model and propose an alternative
strategy relying on variational tools, namely, variational EM.


\subsection{Likelihoods}

Let $\Xbf$ denote the set of all edges, $\Xbf= \{X_{ij}\}_{i, j =
1,\ldots, n}$,
and~$\Zbf$ the set of all indicator variables for nodes, $\Zbf=
\{\Zbf_i\}_{i=1,\ldots,n}$. In the mixture model literature [\cite{MaP00}]
$(\Xbf,
\Zbf)$ is referred to as the complete data set, while $\Xbf$ is
referred to
as the incomplete data set. The conditional independence of the edges
knowing $\Zbf$ entails the decomposition $\log{\Pro(\Zbf,\Xbf)} =
\log{\Pro(\Zbf)} + \log{\Pro(\Xbf|\Zbf)}$. It then follows from
\eqref{Eq:DistZ} and \eqref{Eq:DistX} that the log-likelihood of the
complete data set is
%
%
\begin{equation}
\label{Eq:LXZ}
\log{\Pro(\Zbf,\Xbf)} = \sum_i \sum_q Z_{iq}\log{\alpha_q} +
\sum_{i
\neq j} \sum_{q,\ell} Z_{iq}Z_{j\ell}\log{f_{q\ell}(X_{ij})}.
\end{equation}

The likelihood of the incomplete data set can be obtained by summing
$\Pro(\Zbf,\Xbf)$ over all possible $\Zbf$'s: $\Pro(\Xbf) = \sum
_{\Zbf}
\Pro(\Zbf,\Xbf) $. This summation involves $Q^n$ terms and quickly becomes
intractable. The popular E--M algorithm [\cite{DLR77}], widely used in
mixture problems, allows to maximize $\log{\Pro(\Xbf)}$ without explicitly
calculating it. The E-step relies on the calculation of the conditional
distribution of $\Zbf$ given $\Xbf$: $\Pro(\Zbf|\Xbf)$.
Unfortunately, in
the case of network data, the strong dependency between edges makes this
calculation untractable.



\subsubsection*{Undirected graphs}
The closed formula \eqref{Eq:LXZ} still holds undirected graphs,
replacing the
sum over $i \neq j$ by a sum over $i < j$. This is also true for
equations~\eqref{Eq:JcalRfact} and \eqref{Eq:OptGamma} given below.

\subsection{Variational EM} \label{Subsec:Var}

We propose to use an approximate maximum likelihood strategy based on
a variational approach [see \cite{JGJ99} or the tutorial by
\cite{Jaa00}]. This strategy is also used in \cite{GoN05} for a
biclustering problem. We consider a lower bound of the log-likelihood
of the incomplete data set
%
%
\begin{equation} \label{Eq:DefJ}
\Jcal(\RX,\gammabf) = \log{\Pro(\Xbf;\gammabf)} -
\mathit{KL}(\RX(\cdot),\Pro(\cdot|\Xbf;\gammabf)),
\end{equation}
where $\mathit{KL}$ denotes the Kullback--Leibler divergence and $\RX$ stands for
some distribution on $\Zbf$. Classical properties of the Kullback--Leibler
divergence ensure that~$\Jcal$ has a unique maximum
$\log{\Pro(\Xbf;\gammabf)}$, which is reached for $\RX(\Zbf) =
\Pro(\Zbf|\Xbf)$. In other words, if $\Pro(\Zbf|\Xbf;\gammabf)$ was
tractable, the maximization of $\Jcal(\RX, \gammabf)$ with respect to
$\gammabf$ would be equivalent to the maximization of
$\log{\Pro(\Xbf;\gammabf)}$. In our case, $\Pro(\Zbf|\Xbf
;\gammabf)$ is
untractable and we maximize $\Jcal(\RX,\gammabf)$ with respect to both
$\RX$ and~$\gammabf$. \cite{Jaa00} shows that $\Jcal(\RX,\gammabf
)$ can be
rewritten as
%
%
\begin{equation} \label{Eq:J2}
\Jcal(\RX,\gammabf) = \Hcal(\RX) + \sum_{\Zbf}
\RX(\Zbf)\log{\Pro(\Xbf, \Zbf;\gammabf)},
\end{equation}
where $\Hcal(\cdot)$ denotes the entropy of a distribution. The last term
of \eqref{Eq:J2} can be deduced from \eqref{Eq:LXZ}:
%
%
\begin{eqnarray} \label{Eq:LZR}
& & \sum_{\Zbf} \RX(\Zbf)\log{\Pro(\Xbf, \Zbf;\gammabf)}
\nonumber
\\[-8pt]
\\[-8pt]
& &\quad= \sum_i \sum_q
\Esp_{\RX}(Z_{iq}) \log{\alpha_q} + \sum_{i \neq j} \sum_{q,\ell}
\Esp_{\RX}(Z_{iq} Z_{j\ell}) \log{f_{q\ell}(X_{ij})},\nonumber
\end{eqnarray}
where $\Esp_{\RX}$ denotes the expectation with respect to distribution
$\RX$. Equation (\ref{Eq:LZR}) requires only the knowledge of $\Esp_{\RX}(Z_{iq})$
and $\Esp_{\RX}(Z_{iq} Z_{j\ell})$ for all $i,j,q,\ell$. By contrast,
$\Hcal(\RX)$ requires all order moments of $\RX$ and is untractable in
general. Maximization of $\Jcal(\RX,\gammabf)$ in $\RX$ can not be achieved
without some restrictions on $\RX$. We therefore limit the search to the
class of completely factorized distributions:
%
%
\begin{equation} \label{Eq:Rfact}
\RX(\Zbf) = \prod_i h(\Zbf_i, \taubf_i),
\end{equation}
where $h$ denotes the multinomial distribution and $\taubf_i$ stands
for a
vector of probabilities, $\taubf_i = (\tau_{i1}, \ldots, \tau_{iQ})$ (with
$\sum_q \tau_{iq} = 1$). In particular, $\Esp_{\RX}(Z_{iq}) = \tau_{iq}$
and $\Esp_{\RX}(Z_{iq} Z_{j\ell}) = \tau_{iq} \tau_{j\ell}$. In addition,
the entropy is additive over the coordinates for factorized distributions,
so that $ \Hcal(\RX) = \sum_i \Hcal(h(\cdot,\taubf_i)) = - \sum
_i \sum_q
\tau_{iq} \log{\tau_{iq}}$. Wrapping everything together,
%
%
\begin{eqnarray} \label{Eq:JcalRfact}
\Jcal(\RX,\gammabf) &=& - \sum_i \sum_q \tau_{iq} \log{\tau_{iq}}
+ \sum_i \sum_q \tau_{iq} \log{\alpha_q}\nonumber
\\[-8pt]
\\[-8pt]
&&{} + \sum_{i \neq j}
\sum_{q,\ell} \tau_{iq} \tau_{j\ell} \log{f_{q\ell}(X_{ij})}.\nonumber
\end{eqnarray}

It is immediate from \eqref{Eq:JcalRfact} that $\Jcal(\RX,\gammabf
)$ is
tractable for distributions $\RX$ of the form (\ref{Eq:Rfact}). The
$\taubf_i$'s must be thought of as variational parameters to be
optimized so
that $\RX(\Zbf)$ fits $\Pro(\Zbf|\Xbf;\gammabf)$ as well as
possible; they
depend on the observed data $\Xbf$. Since $\RX$ is restricted to be
of the
form (\ref{Eq:Rfact}), $\Jcal(\RX,\gammabf)$ is a lower bound of
$\log
\Pro(\Xbf)$.

\subsubsection*{Discussion about tighter bounds}

A fully factorized $\RX$ is only one class of distributions we can
consider. Broader distribution classes should yield tighter bound of
$\Jcal(\RX,\gammabf)$. Unfortunately, for more general
distributions, the
entropy $\Hcal(\RX)$ may not have a simple expression anymore
rendering the
exact calculation of $\Jcal(\RX,\gammabf)$ untractable: better
accuracy is
achieved at the expense of tractability. A solution to this issue is Bethe
free energy [\cite{YFW05}]. We did not consider it because it relies
on an
approximation of $\Hcal(\RX)$ which disrupts the well-behaved
properties of
$\Jcal(\RX,\gammabf)$.

Another approach comes from \cite{LeK01} and \cite{Mar06}. Starting
from an
exponential inequality, they emphasize the strong connection between fully
factorized $\RX$ and first order linear approximation of the exponential
function. Using a higher approximation of the exponential and some
distribution $S_{\Xbf}$ in addition to $\RX$, it is possible to
derive an
even tighter bound of $\Jcal(\RX,\gammabf)$. However, the estimation
algorithm is then of complexity $\mathcal{O}(n^6Q^6)$ instead of
$\mathcal{O}(n^2Q^2)$ for a gain which has the same order of magnitude as
the computer numerical precision.

\section{Parameter estimation} \label{Sec:Estim}


We present here the two-steps algorithm used for the parameter estimation.

\subsection{Estimation algorithm}

As explained in Section \ref{Subsec:Var}, the maximum likelihood
estimator of $\gammabf$ is
\[
\widehat{\gammabf}_{\mathit{ML}} =
\operatorname{\arg\max}\limits_{\gammabf} \log{\Pro(\Xbf;\gammabf)} =
\operatorname{\arg\max}\limits_{\gammabf} \max_{\RX} \Jcal(\RX,\gammabf).
\]
In the variational framework, we restrict the last optimization
problem to factorized distributions. The estimate we propose is
hence
\[
\widehat{\gammabf} = \operatorname{\arg\max}\limits_{\gammabf} \max_{\RX
\mathrm{\ factorized}} \Jcal(\RX,\gammabf).
\]
The simultaneous optimization with respect to both $\RX$ and
$\gammabf$ is still too difficult, so we adopt the following
iterative strategy. Denoting by $\RX^{(n)}$ and ${\gammabf}^{(n)}$
the estimates after $n$ steps, we compute
%
%
\begin{equation} \label{Eq:MaxStep}
\cases{
\RX^{(n+1)} = \operatorname{\arg\max}\limits_{\RX\mathrm{\ factorized}}
\mathcal{J}\bigl(\RX,{\gammabf}^{(n)}\bigr), \cr
{\gammabf}^{(n+1)} = \operatorname{\arg\max}\limits_{\gammabf}
\mathcal{J}\bigl(\RX^{(n+1)},\gammabf\bigr).
}
\end{equation}
The next two sections are dedicated to each of these steps.

\subsubsection*{Initialization step}
The optimization procedure \eqref{Eq:MaxStep} only ensures the
convergence toward a local optimum, so the choice of the starting
point for $\gammabf$ or $\RX$ is crucial to avoid local optima. This
choice is difficult, but, to our experience, hierarchical clustering
seems to be a good strategy to get an initial value for $\RX$.

\subsection{Optimal approximate conditional distribution $\RX$}

We consider here the optimization of $\Jcal$ with respect to $\RX$.
For a
given value of $\gammabf$, we\break denote\vspace*{1pt} $\widehat{\taubf}$ the variational
parameter defining the distribution $\wRX=\break {\arg\max}_{\RX\mathrm{\ factorized}}
\Jcal(\RX, \gammabf)$. This amounts to maximimizing $\Jcal(\RX,
\gammabf)$,\break
given in \eqref{Eq:JcalRfact}, under the condition that, for all $i$, the
$\tau_{iq}$'s must sum to 1. The derivative of $\Jcal(\RX, \gammabf
)$ with
respect to $\tau_{iq}$ is
\[
- \log{\tau_{iq}} -1 + \log{\alpha_q} + \sum_{j \neq i}\sum_{\ell}
\tau_{j\ell} [\log{f_{q\ell}(X_{ij})} + \log{f_{\ell
q}(X_{ji})} ] + L_i,
\]
where $L_i$ denotes the $i$th Lagrange multiplier. It results from the
previous equation that the optimal variational parameter $\widehat
{\taubf}$
satisfies the fixed point relation
%
%
\begin{equation} \label{Eq:OptRX}
\widehat{\tau}_{iq} \propto\alpha_q \prod_{j \neq i}
\prod_{\ell} [f_{q\ell}(X_{ij})f_{\ell
q}(X_{ji}) ]^{\widehat{\tau}_{j\ell}}.
\end{equation}

The fixed point relation (\ref{Eq:OptRX}) can be related to a mean field
approximation [see \cite{Jaa00}]. We get $\widehat{\taubf}$ simply by
iterating this relation until convergence.

\subsubsection*{Undirected graphs}
For a undirected graph, $\widehat{\taubf}$ satisfies
\[
\widehat{\tau}_{iq} \propto\alpha_q \prod_{j \neq i} \prod_{\ell}
[f_{q\ell}(X_{ij}) ]^{\widehat{\tau}_{j\ell}}.
\]

\subsection{Parameter estimates}

We now have to maximize $\Jcal$ with respect to $\gammabf= (\alphabf,
\thetabf)$ for a given distribution $\RX$. Again, this amounts to
maximizing $\Jcal(\RX, \gammabf)$, given in \eqref{Eq:JcalRfact}, under
the condition that $\sum_q \alpha_q = 1$. Straightforward
calculations show
that the optimal $\alphabf$ and $\thetabf$ are given by
%
%
\begin{equation} \label{Eq:OptGamma}
\widehat{\alpha}_q = \frac1n \sum_i \tau_{iq},\qquad
\widehat{\theta}_{q\ell} = \operatorname{\arg\max}\limits_{\theta} \sum
_{i \neq
j} \tau_{iq}\tau_{j\ell}\log{f(X_{ij}; \theta)}.
\end{equation}

\subsubsection*{Poisson models} Poisson models are of particular
interest for
our interaction networks. The optimal $\lambda_{q\ell}$ for model PM
presented in Section \ref{subsec:PoisModel} is straightforward:
\[
\widehat{\lambda}_{q\ell} = \sum_{i \neq j} \tau_{iq} \tau_{j\ell}
X_{ij} \Big/ \sum_{i \neq j} \tau_{iq} \tau_{j\ell} .
\]
For models PRMH and PRMI presented in the same section, there is no closed
formula for $\lambda_{q\ell}$, $\beta_{q\ell}$ or $\beta$.
However, since
the Poisson regression model belongs to the exponential family, $\Jcal
$ is
only a weighted version of the log-likelihoods of the corresponding
generalized linear model. As such, standard optimization procedures can be
used.

\subsubsection*{Exponential family} The optimal $\thetabf$ is not
explicit in
the general case, but has a simpler form if the distribution $f$
belongs to
the exponential family. Namely, if $f$ belongs to an exponential family
with natural parameter $\theta$,
\[
f(x; \theta) = \exp{ [ \Psibf(x)' \theta-
A(\theta) ]}.
\]
According to \eqref{Eq:OptGamma}, we look for $\widehat{\theta}=
{\arg\max}_{\theta} \sum_{i \neq j} \tau_{iq}\tau_{j\ell} \Psibf(X_{ij})'
\theta- A(\theta) $. Maximizing this quantity in $\theta$ yields
\[
\sum_{i \neq j} \tau_{iq}\tau_{j\ell} \Psibf(X_{ij}) -
\nabla A(\theta) = \Zerobf.
\]
If $\nabla A$ is invertible, the optimal $\theta$ is
%
%
\begin{equation} \label{Eq:OptThetaEF}
\widehat{\theta} = (\nabla A )^{-1}
\biggl[\sum_{i \neq j} \tau_{iq}\tau_{j\ell} \Psibf(X_{ij}) \biggr].
\end{equation}

\subsection{Choice of the number of groups}

In practice, the number of groups is unknown and should be estimated.
Many criterion have been proposed to select the dimensionality $Q$ of
the latent space, ranging from AIC to ICL. AIC, BIC and their variants
[\cite{BuA98}] are based on computing the likelihood of the observed
data $\Pro(\Xbf|m_Q)$ and penalizing it with some function of~$Q$. But
the use of variational EM is precisely to avoid computation of
$\Pro(\Xbf|m_Q)$, which is untractable. Given a prior distribution
$\Pro(m_Q)$ over models, and a prior distribution $\Pro(\gammabf|m_Q)$
for each model, variational Bayes [\cite{BeG03}] works by selecting
the model with maximum posterior $\Pro(m_Q|\Xbf)$. Estimation of
$\Pro(\Xbf|m_Q)$ is then performed using variational EM and no
penalization is required, as complex models are already penalized by
diffuse prior $\Pro(\gammabf|m_Q)$. Extension of Deviance Information
Criterion (DIC) to finite mixture distributions via variational
approximations [\cite{McGrory2007}] is even more straightforward:
choosing $Q^*$ larger than the expected number of components and
running the algorithm, extraneous classes become void as the
algorithm converges and the selected number of groups is just the
number of nonempty classes. In the context of unknown assignments,
\cite{BCG00} proposed the Integrated Classification Likelihood (ICL),
which is an approximation to the complete data likelihood
$\Pro(\Xbf,\Zbf|m_Q)$. Variational Bayes, BIC and ICL can all be seen
as approximations to Bayes factors. Whereas Variational Bayes
integrates out the uncertainty about the parameter and the assignment
of nodes to groups, ICL replaces them by a point estimate, computed
thanks to variational EM. Traditional model selection essentially
involves a trade-off between goodness of fit and model complexity,
whereas ICL values both goodness of fit and classification sharpness.\looseness=1

\cite{NoS01} do not propose any criterion to select the number of
groups. \cite{HoW08} use McGrory's method but in a very specific case of
the Stochastic Block Model. They also give no clue as to how to decide that
the algorithm has \textit{converged enough}. \cite{ABF08} use either a
modification to BIC (for small size networks) or cross-validation (for
large size networks) to select the number of groups. \cite{DPR08} use a
modification to ICL criterion. Following along the same line as
\cite{DPR08}, we use a modification of ICL adapted to valued graphs to
select the number of classes.

\begin{description}
\item[ICL criterion:]
For a model $m_Q$ with $Q$ classes where $\thetabf$ involves $P_Q$
independent parameters, the ICL criterion is
\begin{eqnarray*}
\mathit{ICL}(m_Q) &=&
\max_{\gammabf} \log\Pro(\Xbf,
\widetilde{\Zbf} |\gammabf, m_Q)\\
&&{} - \frac{1}{2} \{ P_Q \log
[n(n-1)] - (Q-1) \log(n) \},
\end{eqnarray*}
where the missing data $\Zbf$ are replaced by their prediction
$\widetilde{\Zbf}$.
\end{description}


Note that the penalty term $-\frac{1}{2} \{ P_Q \log[n(n-1)] -
(Q-1) \log(n) \}$ is similar to the one of BIC, where the
$\log$ term refers to number of data. In the case of graphs, the
number of data is $n$ (i.e., the number of nodes) for the vector of
proportions $\alphabf$ ($Q-1$ independent parameters), whereas it is
$n(n-1)$ (i.e., the number of edges) for parameter $\thetabf$ ($P_Q$
independent parameters). For the models PM, PRMI and PRMH (detailed in
Section~\ref{subsec:PoisModel}), $P_Q$ is respectively ${Q(Q+1)}/{2}$,
$Q(Q+1)$ and $1 + {Q(Q+1)}/{2}$.

\eject
\section{Simulation study} \label{Sec:Simul}

\subsection{Quality of the estimates} \label{Subsec:EstimQuality}

\subsubsection*{Simulation parameters}
We considered undirected networks of size $n = 100$ and $500$ with
$Q=3$ classes. To study balanced and unbalanced proportions, we set
$\alpha_q \propto a^q$, with $a = 1, 0.5, 0.2$. $a = 1$ gives
uniform proportions, while $a=0.2$ gives very unbalanced
proportions: $\alphabf= (80.6\%, 16.1\%, 3.3\%)$. We finally
considered symmetric connection intensities $\lambda_{pq}$, setting
$\lambda_{pp} = \lambda'$ for all $p$ and $\lambda_{pq} = \lambda'
\gamma$ for $p \neq q$. Parameter $\gamma$ controls the difference
between within class and between class connection intensities
($\gamma= 0.1$, $0.5$, $0.9$, $1.5$), while $\lambda'$ is set so
that the mean connection intensity $\lambda$ ($\lambda= 2, 5$)
depends neither on $\gamma$ nor $a$. $\gamma$ close to one makes the
distinction between the classes difficult. $\gamma$ larger than one
makes the within class connectivities less intense than the between
ones. We expect the fitting to be rather easy for the combination
$\{n=500, a = 1, \lambda=5, \gamma= 0.1\}$ and rather difficult for
$\{n=100, a=0.2, \lambda=2, \gamma=0.9\}$.

\subsubsection*{Simulations and computations}
For each combination of the parameters, we simulated $S = 100$
random graphs according to the corresponding mixture model. We
fitted the parameters using the algorithm described in
Section~\ref{Sec:Estim}. To solve the identifiability problem of the
classes, we systematically ordered them in descending estimated
proportion order: $\widehat{\alpha}_1 \geq\widehat{\alpha}_2 \geq
\widehat{\alpha}_3$. For each parameter, we calculated the estimated
Root Mean Squared Error (RMSE):
\begin{eqnarray*}
\mathit{RMSE}(\widehat{\alpha}_{p}) &=& \sqrt{\frac1S \sum_{s=1}^S
\bigl(\widehat{\alpha_{p}}^{(s)} - \alpha_{p} \bigr)^2},\\
\mathit{RMSE}(\widehat{\lambda}_{pq}) &=& \sqrt{\frac1S \sum_{s=1}^S
\bigl(\widehat{\lambda_{pq}}^{(s)} - \lambda_{pq} \bigr)^2},
\end{eqnarray*}
where the superscript $(s)$ labels the estimates obtained in
simulation $s$. We also calculated the mean posterior entropy
\[
H = \frac1S \sum_s \biggl(- \sum_i \sum_q \tau_{iq}^{(s)} \ln
\tau_{iq}^{(s)} \biggr),
\]
which gives us the degree of uncertainty of the classification.

\subsubsection*{Results}
\begin{figure}

\includegraphics{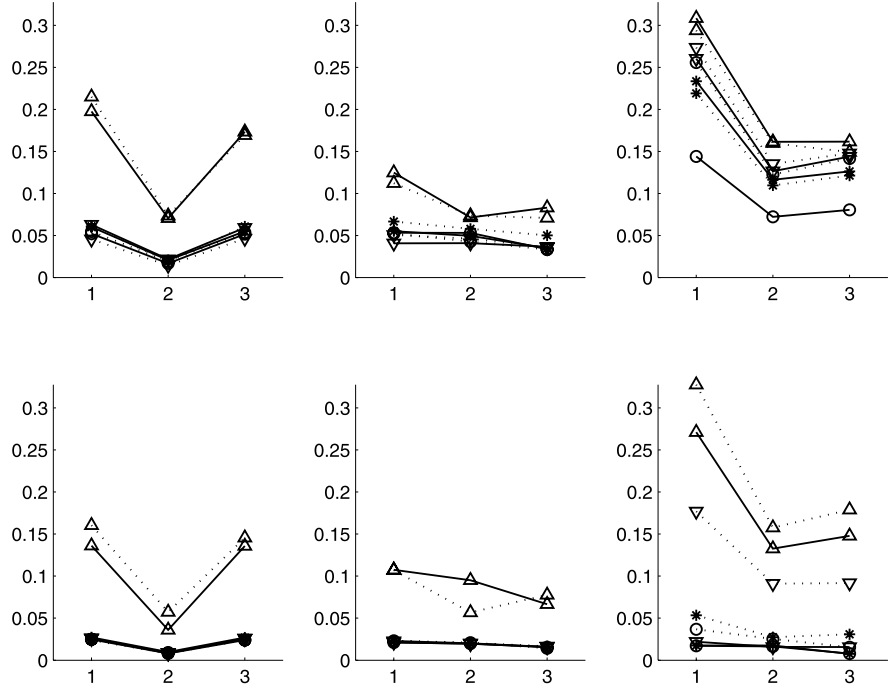}

\caption{$\mathit{RMSE}$ of the estimates $\widehat{\alpha}_q$. The
$x$-axis refers to $\alpha_1, \alpha_2, \alpha_3$. Top: $n = 100$,
bottom: $n = 500$, from left to right: $a=1, 0.5, 0.2$.
Solid line: $\lambda=5$, dashed line: $\lambda= 2$. Symbols
depend on $\gamma$: $\circ=0.1$, $\triangledown=0.5$,
$\vartriangle\,=0.9$, $*=1.5$.} \label{Fig:AlphaRMSE}
\end{figure}

\begin{figure}

\includegraphics{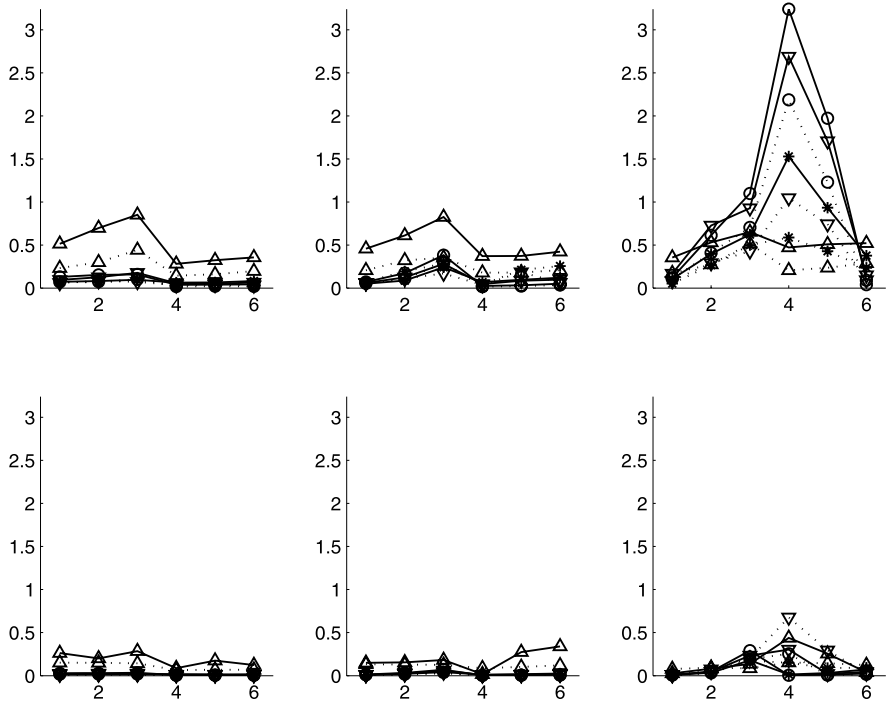}

\caption{$\mathit{RMSE}$ of the estimates $\widehat{\lambda}_{pq}$. The
$x$-axis refers to $\lambda_{11}, \lambda_{22}, \lambda_{33},
\lambda_{12}, \lambda_{13}, \lambda_{23}$. Same legend as
Figure~\protect\ref{Fig:AlphaRMSE}.} \label{Fig:LambdaRMSE}
\end{figure}

Figure~\ref{Fig:AlphaRMSE} (resp. \ref{Fig:LambdaRMSE}) gives
the RMSE for the proportion $\alpha_q$ (resp. connection intensities
$\lambda_{pq}$). As expected, the $\mathit{RMSE}$ is lower when $n$ is
larger. The parameters affecting the $\mathit{RMSE}$ are mainly $a$ and
$\gamma$, whereas $\lambda$ has nearly no effect. The departures
observed for $\alpha_1$ and $\alpha_3$ in the balanced case ($a =
1.0$) are due to the systematic reordering of the proportions.

%

%

Since the graph is undirected, $\lambda_{pq} = \lambda_{qp}$, so
only nonredundant parameters are considered in
Figure~\ref{Fig:LambdaRMSE}. The overall quality of the
estimates is satisfying, especially for the diagonal terms
$\lambda_{qq}$. The within intensity parameter of the smallest class
$\lambda_{33}$ is the most difficult to estimate. The worst case
corresponds to a small graph ($n=100$) with very unbalanced classes
($a = 0.2$) for parameter $\lambda_{12}$. In this case, the
algorithm is unable to distinguish the two larger classes (1 and 2),
so that the estimates extra-diagonal term $\widehat{\lambda}_{12}$
is close to the diagonal ones $\widehat{\lambda}_{11}$ and
$\widehat{\lambda}_{22}$, whereas its true value is up to ten times
smaller.

%
\begin{figure}

\includegraphics{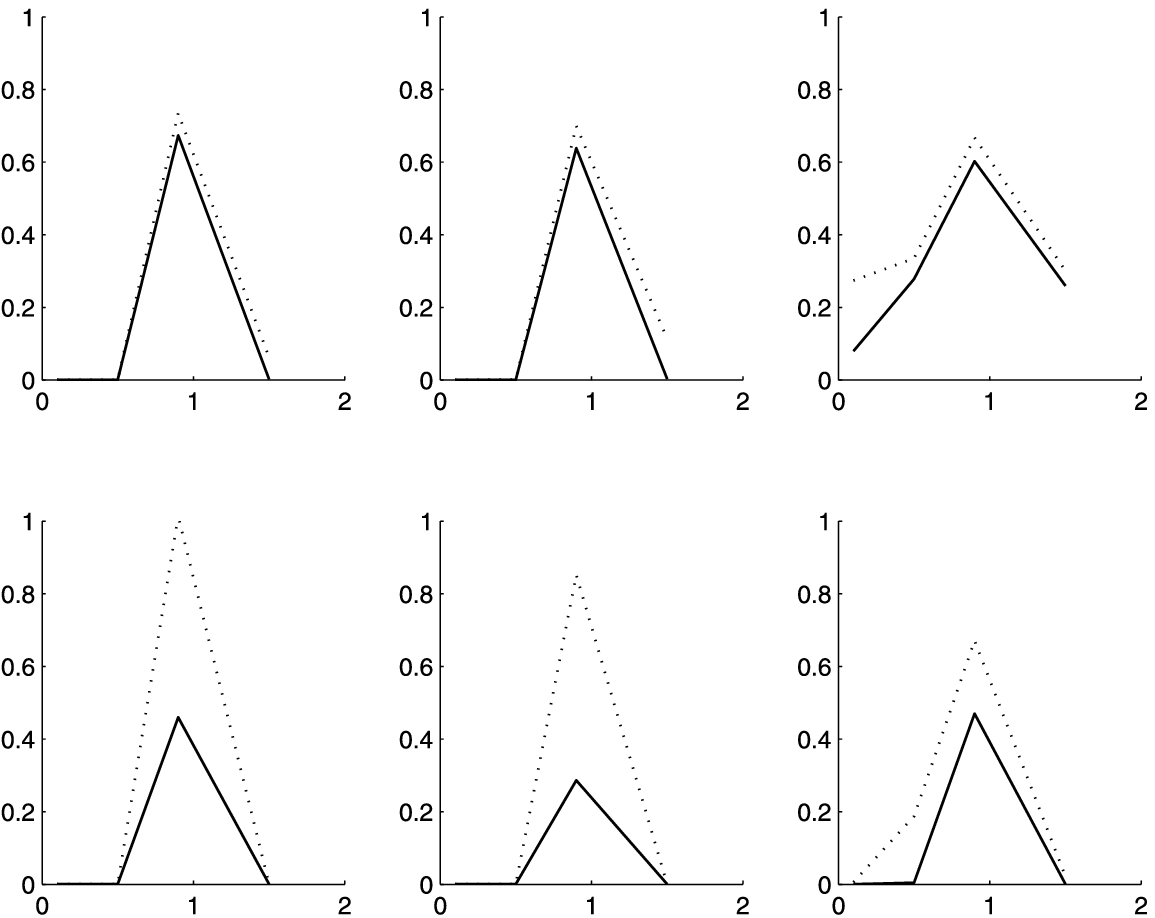}

\caption{Mean (normalized) entropy $H/n$ as a function of
$\gamma$. Top: $n = 100$, bottom: $n = 500$, from left to right:
$a=1, 0.5, 0.2$. Solid line: $\lambda=5$, dashed line: $\lambda= 2$.}
\label{Fig:EntropyMean}
\end{figure}

Figure~\ref{Fig:EntropyMean} gives the mean entropy. Not
surprisingly, the most influential parameter is $\gamma$: when $\gamma$
is close to 1, the classes are almost indistinguishable. For small
graphs ($n=100$), the mean intensity $\lambda$ has almost no effect.
Because of the identifiability problem already mentioned, we did not
consider the classification error rate.

\subsection{Model selection}

We considered a undirected graph of size $n=50,\break 100,  500$ and
$1000$ with $Q^\star=3$ classes. We considered the combination $\{a
= 0.5, \lambda=2, \gamma= 0.5\}$ which turned out to be a medium
case (see Section~\ref{Subsec:EstimQuality}) and computed ICL for
$Q$ ranging from $1$ to $10$ (from $1$ to $5$ for $n=1000$) before
selecting the $Q$ maximizing ICL. We repeated this for $S=100$
simulations.

%
%
%
\begin{figure}
\centering
\begin{tabular}{cc}

\includegraphics{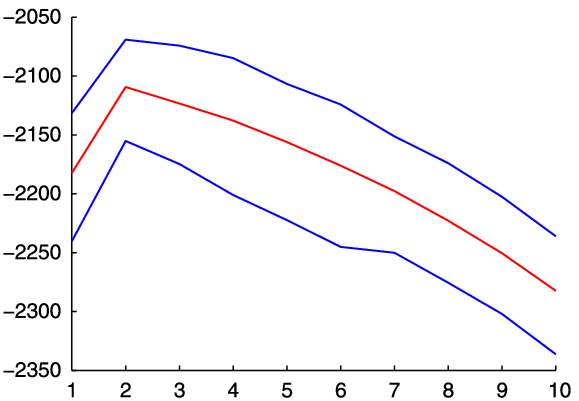}
&\includegraphics{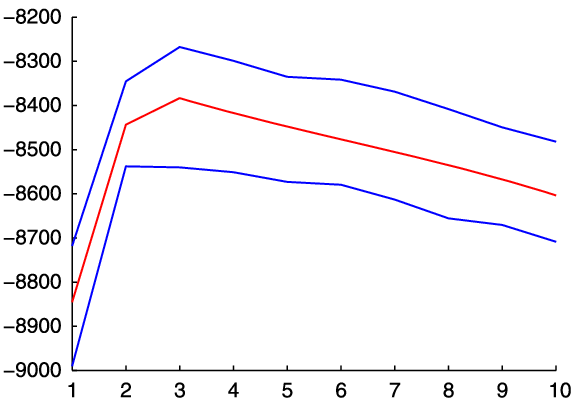}\\
\footnotesize{(a)}&\footnotesize{(b)}\\[6pt]

\includegraphics{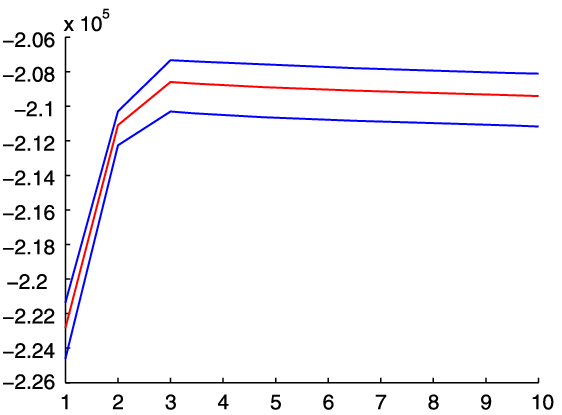}
&\includegraphics{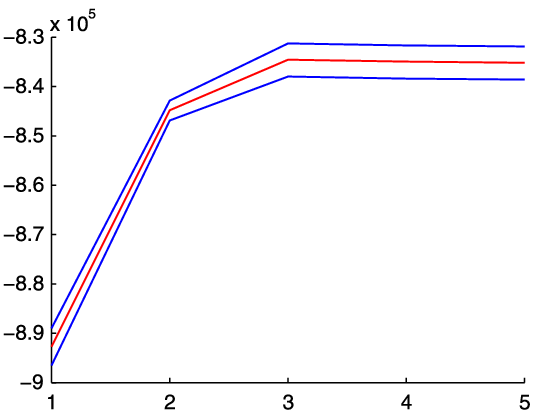}\\
\footnotesize{(c)}&\footnotesize{(d)}
\end{tabular}
%
%
\caption{Mean ICL and 90\% confidence interval as a function of
$Q$. (\textup{a}) $n=50$, (\textup{b}) $n=100$, (\textup{c})~$n=500$, (\textup{d}) $n=1000$.} \label{Fig:ICL}
\end{figure}

Figure~\ref{Fig:ICL} gives ICL as a function of $Q$, while
Table~\ref{Tab:SelectedQ} returns the frequency with which each
$Q$ is selected. As soon as $n$ is larger than $100$, ICL almost
always selects the correct number of classes; for smaller graphs
($n=50$), it tends to underestimate it. The proposed criterion is
thus highly efficient.

\section{Uncovering the structure of host--parasite interactions in forest
ecosystems} \label{Sec:Ecology}

Here we use mixture models to highlight the factors governing with
whom a species interact in an ecosystem. The factors which may account
for species interactions are introduced as covariates in the mixture
models. The explanatory power of each factor is measured as the
decrease in the number of groups selected. Our study focuses on
host--parasite interactions in forest ecosystems. We address the two
following questions: (1) Is similarity in the parasite assemblages of
two tree species explained by their phylogenetic relatedness rather
than by the degree of overlap of their distributional range? (2) Is
similarity in the host range of two parasitic fungal species explained
by their phylogenetic relatedness rather than their common nutritional
strategy? The explanatory power of phylogenetic relatedness is
subsequently called phylogenetic signal, as in the ecological
literature [\cite{RLG07}; \cite{VPD08}].

\subsection{Data}

\subsubsection*{Host--parasite interaction records}
We considered two undirected, valued networks having parasitic fungal
species ($n=154$) and tree species ($n=51$) as nodes, respectively. Edges
strength was defined as the number of shared host species and the
number of
shared parasitic species, respectively [\cite{MRV10}].

The methods used for collecting data on tree--fungus interactions are
fully described in \cite{VPD08}. Fungal species names were checked
since then in the Index Fungorum database
(\href{http://www.indexfungorum.org}{www.indexfungorum.}
\href{http://www.indexfungorum.org}{org}): 17 names
were updated, yielding to 3
new species synonymies. The fusion of synonym species accounts for the
lower number of fungal species in the present study than in the
original publication.

%
\begin{table}[b]
\caption{Frequency (in \%) at which $Q$ is
selected for various
sizes $n$}\label{Tab:SelectedQ}
%
\begin{tabular*}{212pt}{@{\extracolsep{\fill}}ld{2.0}d{2.0}d{3.0}d{3.0}@{}}
\hline
  & \multicolumn{4}{c@{}}{\textit{\textbf n}} \\[-5pt]
   & \multicolumn{4}{c@{}}{\hrulefill} \\
$\mathbf{Q}$ & \multicolumn{1}{c}{\textbf{50}} & \multicolumn{1}{c}{\textbf{100}} & \multicolumn{1}{c}{\textbf{500}} & \multicolumn{1}{c@{}}{\textbf{1000}} \\
\hline
2 & 82 & 7 & 0 & 0 \\
3 & 17 & 90 & 100 & 100 \\
4 & 1 & 3 & 0 & 0\\
 \hline
\end{tabular*}
%
\end{table}

\subsubsection*{Phylogenetic relatedness between species}
In order to verify the existence of a phylogenetic signal in the
parasite assemblages of tree species, we estimated genetic distances
between all pairs of tree species. The maximally resolved seed plant
tree of the software Phylomatic2 [\cite{WeD05}] was used to produce a
phylogenetic tree for the 51 tree species included in our study. Then,
pairwise genetic distances (in million years) were extracted by using
the cophenetic.phylo function of the R ape package [\cite{PCS05}].
Because the phylogenetic tree was loosely resolved for gymnosperms, we
also used taxonomic distances to estimate phylogenetic relatedness
between tree species.
Since all tree species included in the study belong to the phylum
Streptophyta, we used the finer taxonomic ranks of class, order,
family and genus to calculate pairwise taxonomic distances. Based on
the NCBI Taxonomy Browser
(\href
{http://www.ncbi.nlm.nih.gov/Taxonomy/}{www.ncbi.nlm.nih.gov/Taxonomy/}),
we found that the
species are evenly distributed into two taxonomic classes
(Magnoliophyta and Conipherophyta) and further subdivided in 8 orders,
13 families and 26 genera. Following \cite{Pou05}, we considered that
the taxonomic distance is equal to $0$ if species are the same, $1$ if
they belong to the same genus, $2$ to the same family, $3$ to the same
order, $4$ to the same taxonomic class and $5$ if their only common
point lies in belonging to the phylum Streptophyta.

In order to investigate the existence of a phylogenetic signal in the
host range of parasitic fungal species, we estimated taxonomic
distances between all pairs of fungal species. Pairwise genetic
distances could not be calculated because genetic data were not
available for all the species. Since the 153 fungal species at hand
span a wider portion of the tree of life than the tree species, we had
to use the higher order rank of kingdom. The taxonomic distance for
fungal species thus ranges from $0$ to $6$ (kingdom level) when
compared to $0$ to $5$ for trees. The taxonomy was retrieved from
Index Fungorum (\href
{http://www.indexfungorum.org}{www.indexfungorum.org}). All fungal species
included in the study belong to the Fungi kingdom, are divided in two
phyla (Ascomycota and Basidiomycota) and further subdivided in 9
taxonomic classes, 21 orders, 48 families and 107 genera. When pairs
included a species whose taxonomic is uncertain for a given taxonomic
rank, this rank was skipped and upper ranks were used to estimate
distance.

\subsubsection*{Other explanatory factors}
Other factors than phylogenetic relatedness may account for pairwise
similarities in parasite assemblages between tree species. In
particular, two tree species having overlapping distributional range
are exposed to similar pools of parasitic species and may therefore
share more parasitic species than two tree species with
nonoverlapping distributions [\cite{BrB06}]. We tested this
hypothesis by calculating the geographical distance between all pairs
of tree species. The geographical distance is the Jaccard distance
[\cite{Jac01}] computed on the profiles of presence/absence in 309
geographical units covering the entire French territory.

%
\begin{table}[b]
\tablewidth=278pt
\caption{Top: Size, mean number of interactions and Magnoliophyta content
for each group found with PM. Bottom: Parameter
estimates for the tree network: $\lambda_{q\ell}=$ mean number of
shared parasitic
species, $\alpha_q = $ group proportion (\%) with PM (no covariate)}
\label{Tab:TreePM}

\includegraphics{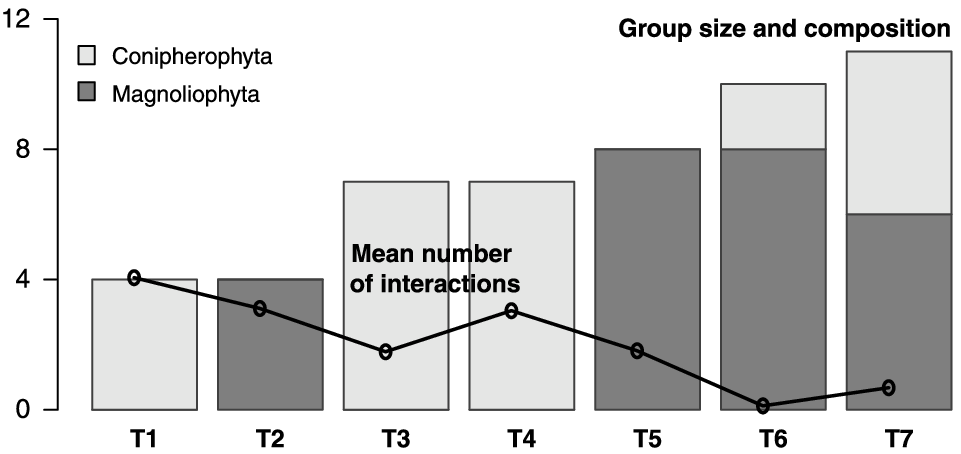}

\begin{tabular*}{278pt}{@{\extracolsep{\fill}}ld{2.2}d{2.2}d{2.2}d{2.2}d{2.2}d{2.2}d{2.2}@{}}
\hline
$\bolds{\widehat{\lambda}_{\bolds{q}\ell}}$ & \multicolumn{1}{c}{\textbf{T1}} & \multicolumn{1}{c}{\textbf{T2}}
& \multicolumn{1}{c}{\textbf{T3}} & \multicolumn{1}{c}{\textbf{T4}}
& \multicolumn{1}{c}{\textbf{T5}} & \multicolumn{1}{c}{\textbf{T6}} &\multicolumn{1}{c@{}}{\textbf{T7}} \\
\hline
T1 & 14.46 & 4.19 & 5.99 & 7.67 & 2.44 & 0.13 & 1.43 \\
T2 & 4.19 & 14.13 & 0.68 & 2.79 & 4.84 & 0.53 & 1.54 \\
T3 & 5.99 & 0.68 & 3.19 & 4.10 & 0.66 & 0.02 & 0.69 \\
T4 & 7.67 & 2.79 & 4.10 & 7.42 & 2.57 & 0.04 & 1.05 \\
T5 & 2.44 & 4.84 & 0.66 & 2.57 & 3.64 & 0.23 & 0.83 \\
T6 & 0.13 & 0.53 & 0.02 & 0.04 & 0.23 & 0.04 & 0.06 \\
T7 & 1.43 & 1.54 & 0.69 & 1.05 & 0.83 & 0.06 & 0.27 \\
[6pt]
$\widehat{\alpha}_q$ & 7.8 & 7.8 & 13.7 & 13.7 & 15.7 & 19.6 &
21.6\\
\hline
\end{tabular*}
\end{table}

In the case of fungal species, other factors may also account for
similarity in host range. Here we investigated whether fungal species
having similar nutritional strategies also have similar host ranges. Fungal
species were classified into ten nutritional strategies based on their
parasitic lifestyle (biotroph or necrotroph) and on the plant organs and
tissues attacked. Five strategies (strict foliar necrotroph parasites,
canker agents, stem decay fungi, obligate biotroph parasites and root decay
fungi) accounted for 87\% of the fungal species. We considered that
nutritional distance between two species equals one if the strategies are
the same and 0 otherwise.

\subsection{Identification of groups of species sharing similar
interactions}

\subsubsection*{Model}
For both networks, we used the mixture model to define groups of tree
species and fungal species having similar interactions We assumed
that, in each network, the edge intensities were Poisson distributed.
For both networks, we considered the PM and PRMH models (see Section
\ref{subsec:PoisModel}) using pairwise distance between species
(genetic, taxonomic, geographic or nutritional) as a covariate.




\subsubsection*{PM model: No covariate}
In the absence of covariates, the ICL criterion selected 7 groups of
tree species. Two groups of tree species (T2 and T5) were exclusively
composed of species belonging to the Magnoliophyta, whereas three other
groups (T1, T3 and T4) were exclusively composed of species belonging
to the Conipherophyta. The two last groups (T6 and T7) were mixed
(Table~\ref{Tab:TreePM}). According to the mean number of interactions
per species and the parameters estimates of the model
(Table~\ref{Tab:TreePM}), they were composed of tree species having few
parasitic species and sharing few of them with other tree species.

It is noteworthy that group T2 was composed of four species belonging
to the same order (Fagales) and also to the same family (Fagaceae).
Groups T1, T3 and T4 were also composed of species belonging to the
same family (Pinaceae) since the only three coniferous species
belonging to another family were classified in groups T6 and T7. These
results confirm that two plant species with a similar evolutionary
history are likely to share the same set of parasitic species
[\cite{BrB06}, \cite{GiW07}, \cite{VPD08}].

\subsubsection*{PRMH model: Accounting for phylogenetic relatedness}
When accounting for taxonomy, ICL selected only $4$ groups of tree
species. The estimated regression coefficient was $\widehat{\beta} = -0.317$,
which means that, for the mean taxonomic distance $\overline{y} = 3.82$,
the mean connexion intensity is reduced of $70\%$
($e^{\widehat{\beta}\overline{y}} = 0.298$). The cross classification table
(Table~\ref{Tab:TreeCross}) shows that the taxonomic distance reduces the
number of class by merging groups T1 and T2 with most of the trees of T4
and T5. T'3 essentially consists of T6, T'1 of T7 and T'2 is made of trees
from T3 completed with leftovers from other classes. Interestingly and
unlike the groups obtained with no covariates, no group has species
belonging exclusively to one or the other of the taxonomic classes
(Magnoliophyta or Conipherophyta): the association between group of trees
and taxonomy was cropped out by the covariate
(Table~\ref{Tab:TreePRMH}). The same results hold when using the genetic
distance as a covariate instead of the taxonomic distance (results not
shown).

%
\begin{table}[b]
\tablewidth=240pt
\caption{Cross classification of the groups of tree selected found by
PM and PRMH (with taxonomic variate as a covariate)} \label{Tab:TreeCross}
\begin{tabular*}{240pt}{@{\extracolsep{\fill}}lccd{2.0}c@{}}
\hline
& \textbf{T'1} & \textbf{T'2} & \multicolumn{1}{c}{\textbf{T'3}} & \textbf{T'4} \\
\hline
T1 & 0 & 0 & 0 & 4 \\
T2 & 0 & 0 & 0 & 4 \\
T3 & 2 & 5 & 0 & 0 \\
T4 & 0 & 2 & 0 & 5 \\
T5 & 0 & 2 & 0 & 6 \\
T6 & 0 & 0 & 10 & 0 \\
T7 & 7 & 2 & 2 & 0\\
\hline
\end{tabular*}
\end{table}


Therefore, the inclusion of taxonomic (or genetic) distance as a
covariate shows that the phylogenetic relatedness between tree species
accounts for a large part of the structure of tree--parasitic fungus
interactions in forest ecosystems, but not for all the structure.
Indeed, even after controlling for the evolutionary history through
the taxonomic (or genetic) distance, ICL still finds $4$ groups of
trees, whereas we would expect only one group if the phylogeny was the
sole source of structure. Below we investigate whether the
distributional overlap between tree species is another source of
structure.
%
\begin{table}
\tablewidth=282pt
\caption{Top: Size, mean number of interactions (scaled by $\times$3)
and Magnoliophyta content for each group found with PRMH. Bottom: Parameter
estimates for the tree network: $\lambda_{q\ell}=$ mean number of
shared parasitic
species, $\alpha_q = $ group proportion (\%) with PRMH (with
covariate), $\widehat{\beta} =$ covariate regression coefficient}
\label{Tab:TreePRMH}

\includegraphics{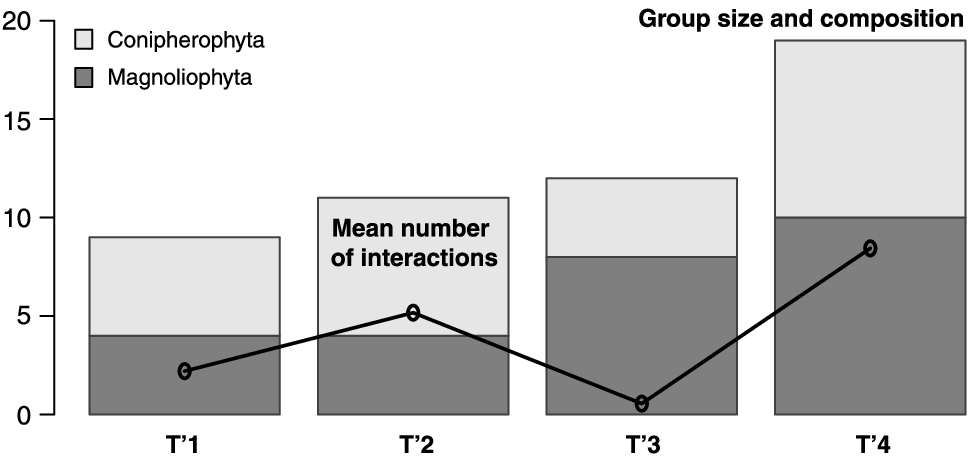}

\begin{tabular*}{282pt}{@{\extracolsep{\fill}}ld{2.2}d{2.2}d{2.3}d{2.2}@{}}
\hline
$\bolds{\widehat{\lambda}_{q\ell}}$ & \multicolumn{1}{c}{\textbf{T'1}} & \multicolumn{1}{c}{\textbf{T'2}}
& \multicolumn{1}{c}{\textbf{T'3}} & \multicolumn{1}{c@{}}{\textbf{T'4}} \\
\hline
T'1 & 0.75 & 2.46 & 0.40 & 3.77 \\
T'2 & 2.46 & 4.30 & 0.52 & 8.77 \\
T'3 & 0.40 & 0.52 & 0.080 & 1.05 \\
T'4 & 3.77 & 8.77 & 1.05 & 14.22 \\[6pt]
$\widehat{\alpha}_q$ & 17.7 & 21.5 & 23.5 & 37.3 \\[6pt]
$\widehat{\beta}$ & \multicolumn{4}{c@{}}{$-0.317$}\\
\hline
\end{tabular*}
\end{table}

\subsubsection*{PRMH model: Accounting for distributional overlap}
In contrast with the taxonomic and genetic distance, the geographical
distance between species does not reduce the number of groups (not
shown). This result suggests that the current distributional overlap
between tree species does not account for the similarity in their
parasite assemblages. This result is opposite to the conventional
wisdom in the field of community ecology, which favors ecological
processes, taking place over short time scale, over evolutionary
processes, taking place over longer time scales, as the main source of
biotic interaction diversity. Our findings point out that the relative
importance of these processes might be the other way round.

\subsection{Factors accounting for the host ranges of parasitic fungal
species}

\subsubsection*{PM model: No covariate}
The ICL criterion selected 9 groups of parasitic fungal species. The
estimates intensities $\widehat{\lambda}_{q\ell}$ range from almost zero
($1.4\times 10^{-3}$) to $12.1$, while the group proportions
$\widehat{\alpha}_q$ range from $1.3\%$ to $40.2\%$ (Table~\ref
{Tab:FungusPM}).

\subsubsection*{PRMH model: Accounting for phylogenetic relatedness}
Accounting for taxonomic distance does not reduce the number of groups
(not shown), indicating a lack of phylogenetic signal in the host
range of fungal species. These results parallel those obtained with
another clustering approach [\cite{New04}] for the same tree--fungus
network [\cite{VPD08}]. They are congruent with the results obtained
for other bipartite networks since asymmetries in the phylogenetic
signal have been found in numerous plant--animal mutualistic networks
[\cite{RLG07}] and in a host--parasite network between leaf-miner moths
and parasitoid insects [\cite{IvG06}]. In the latter case, the authors
also observed a lack of signal through the parasite phylogeny. In the
case of the tree--fungus network, we proposed that the very early
divergence of the major fungal phyla may account for the asymmetric
influence of past evolutionary history [\cite{VPD08}]: the lack of
signal through the fungal phylogeny may be the result of parasitic
fungal species splitting into two groups when the Conipherophyta and
the Magnoliophyta diverged (both groups containing Ascomycota and
Basidiomycota species) and the subsequent coevolution of each set of
fungal species with its plant phylum. Stronger selection pressures on
parasitic species than on host species might also account for the
asymmetry of the signal [\cite{BeK08}; \cite{RMA06}].

\subsubsection*{PRMH model: Accounting for nutritional strategies}
Fungal
Correlation analysis showed an association between the 9 groups
selected with the PM model and the nutritional type.
In particular, two groups of fungal species (F2 and F3, see Appendix
\ref{Sec:ResFungi}) contained a high proportion of root decay fungi
(100\% and 75\%, respectively).
However, taking the nutritional strategy as a covariate does not
reduce the number of groups, indicating the lack of `nutritional
signal' in the host range of parasitic fungal species.

\subsection{Goodness of fit}
Since no covariate decreases the number of mixture components in the
fungus interaction network, we assessed goodness of fit only for the
tree interaction network. The goodness is assessed in two ways: in
terms of likelihood with the ICL criterion and in terms of predictive
power for the strength of an interaction. The ICL criterion is
$-2876.6$ for the base model with no class. It jumps to $-1565.6$
($\Delta \mathit{ICL} = 1212.8$) when allowing a mixture structure (with $7$
classes). It jumps again to $-1449.6$ ($\Delta \mathit{ICL} = 116$) when adding
the taxonomic distance as a covariate in the model (with $4$ classes).
Interestingly, adding a covariate to the $4$ class mixture model
provides a gain in goodness of fit twice as big as the gain of adding
three additional classes ($\Delta \mathit{ICL} = 214.2$ against $98.2$). But
adding a covariate only requires one additional parameter ($\beta$),
against $21$ for the three additional classes.

%
\begin{figure}

\includegraphics{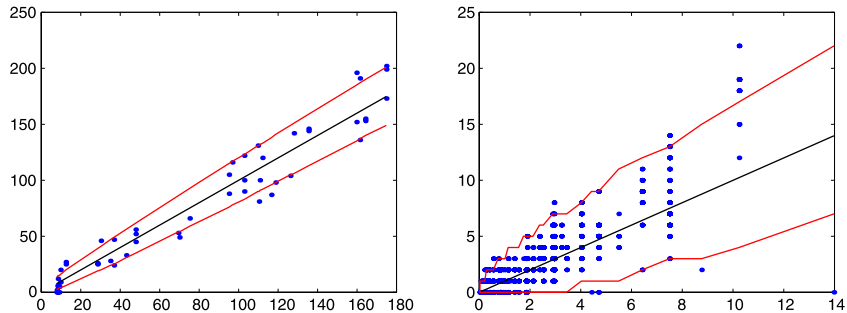}

\caption{Left: Observed versus predicted graph of
the weighted degree $K_i$ of node $i$ ($R^2 = 0.94$). Right: Observed
versus predicted graph of single edge values $X_{ij}$ ($R^2=0.56$).
Black: regression line; red: Poisson 95\% confidence interval.}\label
{fig:goodoffit}
\end{figure}
%
\begin{table}[b]
\caption{Tree interaction network. Effect of different factors on the similarity in parasite
assemblages between tree species. $\Delta $ICL is the gain (in
log-likelihood units) obtained when switching from the best PM model to
the best PRMH model for a given covariate} \label{Tab:Conc}
%
\begin{tabular*}{\textwidth}{@{\extracolsep{\fill}}lcccd{3.1}@{}}
\hline
\textbf{Factor} & \textbf{Covariate} & \textbf{Nb. groups (PM)} &
\textbf{Nb. groups (PRMH)} & \multicolumn{1}{c@{}}{$\bolds\Delta\bolds{\mathit{ICL}}$}\\
\hline
Phylogenetic & Taxon. dist. & 7 & 4 & 116.0 \\
\quad relatedness & Genetic dist. & 7 & 4 & 94.8 \\
[3pt]
Distributional & Jaccard dist. & 7 & 7 & -8.6 \\
\quad overlap \\
\hline
\end{tabular*}
%
\end{table}

We also assessed goodness of fit in terms of predictive power. For the PRMH
model with 4 classes and taxonomic distance as a covariate, we can predict
both the weighted degree $K_i = \sum_{j \neq i} X_{ij}$ of node $i$ as
$\widehat{K}_i = \sum_j \sum_{q,\ell} \tau_{iq} \tau_{jl}
\lambda_{ql}e^{\beta^{\intercal}\ybf_{ij}}$ and the value $X_{ij}$
of a
single edge fungal as $\widehat{X}_{ij} = \sum_{q,\ell} \tau_{iq}
\tau_{j\ell}
\lambda_{ql} e^{\beta^{\intercal}\ybf_{ij}}$. The prediction of
$K_i$ using
$\widehat{K}_i$ is pretty accurate (Figure~\ref{fig:goodoffit} left,
$R^2=0.94$). The prediction of $X_{ij}$ using $\widehat{X}_{ij}$ is less
accurate, but the confidence region is still pretty good
(Figure~\ref{fig:goodoffit} right, $R^2=0.56$).

\subsection{Conclusion}

The structure of host--parasite interactions in forest ecosystems is a
complex one. Some tree species share more parasites than others and
this variability is well captured by a mixture model. However and as
shown in Table~\ref{Tab:Conc}, the naive mixture model deceptively
captures part of the variability readily explained by other factors,
such as the phylogenetic relatedness (measured either by taxonomic or
genetic distance) and artificially increases the number of groups in
the mixture. Accounting for relevant factors decreases the number of
groups selected.
Using group reduction as a yardstick (Table~\ref{Tab:Conc}), we
conclude that similarity in the parasite assemblages of tree species
is explained by their phylogenetic relatedness rather than their
distributional overlap, indicating the importance of evolutionary
processes for explaining the current patterns of inter-specific
interactions. Our study is however inconclusive on the relative
contribution of phylogenetic relatedness and nutritional strategy to
the similarity in the host ranges of parasitic fungal species
parasites of two parasitic fungus (Table~\ref{Tab:Conc2} in Appendix
\ref{Sec:ResFungi}). In either case, since the PRMH model still finds
$4$ (resp. $9$) classes for the tree species (resp. fungal species)
interaction network, a significant fraction of the variability remains
unexplained by our predictors.

\begin{appendix}

\section*{Appendix}

\subsection{Other mixture models} \label{SubSec:ClassicDist}

We examine here some other classical distributions which can be used
in our framework.
\begin{description}
\item[Bernoulli.] In some situations such as co-authorship or
social networks, the only available information is the presence or
absence of the edge. $X_{ij}$ is then supposed to be Bernoulli
distributed:
\[
X_{ij}|i\in q, j \in\ell\sim\Bcal(\pi_{q\ell}).
\]
It is equivalent to the stochastic block model of \cite{NoS01} or
\cite{DPR08}.
%
\item[Multinomial.] In a social network, $X_{ij}$ may specify the
nature of the relationship: colleague, family, friend, etc.
The $X_{ij}$'s can then be modeled by multinomial variables:
\[
X_{ij}| i \in q, j \in\ell\sim\Mcal(1; \pbf_{q\ell}).
\]
The parameter $\theta_{q\ell}$ to estimate is the vector of
probability $\pbf_{q\ell} = (p_{q\ell}^1, \dots,
p_{q\ell}^m )$, $m$ being the number of possible labels.

\begin{table}[b]
\caption{Estimates of $\theta_{q\ell}$ for
some classical distributions. Notation is defined in Section
\protect\ref{SubSec:ClassicDist}. $\kappa_{q\ell}$ stands for $1/\sum_{i
\neq j}\tau_{iq} \tau_{j\ell}$. $\Wbf_{q\ell}$ is the diagonal
matrix with diagonal term $\tau_{iq}\tau_{j\ell}$. \# param. is
the number of independent parameters in the case on directed
graph, except for the bivariate Gaussian only defined for a
nonoriented graph}\label{Tab:ClassicTheta}
\begin{tabular*}{\textwidth}{@{\extracolsep{\fill}}lll@{}}
\hline
\textbf{Distribution}  & \multicolumn{1}{c}{\textbf{Estimate}}  & \multicolumn{1}{c@{}}{\textbf{\# param.}}  \\
\hline
\mbox{Bernoulli} & $\widehat{\pi}_{q\ell} ={\kappa_{q\ell}\sum_{i \neq j} \tau_{iq} \tau_{j\ell} X_{ij}}$ & $Q^2$ \\[1pt]
\mbox{Multinomial} & $\widehat{p}^k_{q\ell} =
{\kappa_{q\ell} \sum_{i \neq j} \tau_{iq}
\tau_{j\ell} \Ibb(X_{ij} = k)}$ & $(m-1)Q^2$ \\[1pt]
\mbox{Gaussian} & $\widehat{\sigma}^2_{q\ell} =
{\kappa_{q\ell} \sum_{i \neq j} \tau_{iq}
\tau_{j\ell} (X_{ij} - \widehat{\mu}_{q\ell})^2}$ & $Q^2$ \\
\mbox{Bivariate Gaussian} & $\widehat{\mubf}_{q\ell} =
{\kappa_{q\ell} \sum_{i \neq j} \tau_{iq}
\tau_{j\ell} \Xbf_{ij}}$ & $Q(Q+1)$ \\
& $\widehat{\Sigmabf}_{q\ell} = {\kappa_{q\ell}
\sum_{i \neq j} \tau_{iq} \tau_{j\ell} (\Xbf_{ij} -
\widehat{\mubf}_{q\ell} ) (\Xbf_{ij} -
\widehat{\mubf}_{q\ell} )'}$ & ${\frac32 Q(Q+1)}$ \\
\mbox{Linear regression} & $\widehat{\betabf}_{q\ell} =
{ (\Ybf' \Wbf_{q\ell}^{-1} \Ybf )^{-1}
\Ybf' \Wbf_{q\ell}^{-1} \Xbf}$ & $pQ^2$\\
& $\widehat{\sigma}^2_{q\ell} = {\kappa_{q\ell}
\sum_{i \neq j} \tau_{iq} \tau_{j\ell} (X_{ij} -
\ybf'_{ij} \widehat{\beta}_{q\ell} )^2}$ & $Q^2$\\
\mbox{Simple regression} & $\widehat{b} = \frac{
\sum_{i \neq j} \sum_{q,\ell}
\tau_{iq} \tau_{j\ell} (X_{ij} -
\overline{X}_{ql} ) (y_{ij}-\bar{y}_{ql} )}{
\sum_{i \neq j} \sum_{q,\ell}\tau_{iq} \tau_{j\ell}
(y_{ij}-\overline{y}_{ql} )^2}$ & 1 \\
& $\widehat{\alpha}_{ql} = \overline{X}_{q\ell} - \widehat{b}\overline
{y}_{ql}$ & $Q^2$
\\
& $\widehat{\sigma}^2 = \frac{1}{n} \sum_{i \neq j} \sum
_{q,\ell}
\tau_{iq} \tau_{j\ell} (X_{ij}- \widehat{\alpha}_{q\ell} y_{ij})^2$ & 1\\
\hline
\end{tabular*}
\end{table}

In directed random graphs, this setting allows to account for some
dependency between symmetric edges $X_{ij}$ and $X_{ji}$. We only need
to consider the equivalent undirected graphs where edge
$(i, j)$ is labeled with the couple $(X_{ij}, X_{ji})$. $m=4$
different labels can the be observed: $(0, 0)$ if no edge exists,
$(1, 0)$ for $i \rightarrow j$, $(1, 1)$ for $i \leftarrow j$ and
$(1, 1)$ for $i \leftrightarrow j$.
%
\item[Gaussian.] Traffic networks describe the intensity of
the traffic between nodes. The airport network is a typical example
where the edges are valued according to the number of passengers
traveling from airport $i$ to airport $j$. The intensity $X_{ij}$
of the traffic can be assumed to be Gaussian:
\[
X_{ij}|i \in q, j \in\ell\sim\Ncal(\mu_{q\ell},\sigma^2_{q\ell}),\qquad
\theta_{q\ell} = (\mu_{q\ell}, \sigma^2_{q\ell}).
\]
%
%
\item[Bivariate Gaussian.] The correlation between symmetric edges
$X_{ij}$ and $X_{ji}$ can be accounted for, considering the
undirected valued graph where edge $(i, j)$ is valued by $(X_{ij},
X_{ji})$, which is assumed to be Gaussian. Denoting $\Xbf_{ij} =
[X_{ij} X_{ji}]'$,
\[
\Xbf_{ij} | i \in q, j \in\ell\sim\Ncal(\mubf_{q\ell},
\Sigmabf_{q\ell}),\qquad
\theta_{q\ell} = (\mubf_{q\ell}, \Sigmabf_{q\ell}).
\]
%
%
\item[Linear regression.] When covariates are available, the
linear model, either Gauss\-ian for real valued edges or generalized for
integer valued (e.g., Poisson or Bernoulli) allows to include them. For
example, for Gaussian valued edges, denoting $\ybf_{ij}$ the $p \times1$
vector of covariates describing edge $(i, j)$, we set
\[
X_{ij}|i \in
q, j \in\ell\sim\mathcal{N}({\bf\beta}_{q\ell}^{\intercal}.\mathbf
{y}_{ij},\sigma_{q\ell}^2).
\]
\item[Simple linear regression.] A case of specific interest for
plant ecology is the simple linear homoskedastic regression with group
specific intercept $a_{q\ell}$ but constant regression coefficient
$b$. It is
particularly useful when controlling for the effect of geography, which
is assumed to be the same for all groups of plants. We then set
\[
X_{ij}| i \in q, j \in\ell\sim\mathcal{N}(a_{q\ell} + b y_{ij},
\sigma^2).
\]
The model can again be extended to Poisson or Bernoulli valued edges
using adequate link function.
\end{description}



\subsection{Parameter estimates for other distributions}

Table \ref{Tab:ClassicTheta} gives the parameter estimates for the
model listed in Section \ref{SubSec:ClassicDist}. The estimates of the
mean parameter for Gaussian ($\mu_{q\ell}$) distributions are the same
as the estimate of the probability $\pi_{q\ell}$ in the Bernoulli
case. The results displayed in this table are all straightforward.
Note that all estimates are weighted versions of the intuitive ones.\looseness=1

\subsection{Parameter estimates for the fungus interaction network}
\label{Sec:ResFungi}

\mbox{}



\begin{table}[h]
\caption{Top: Size, mean number of interactions ($\bar{\lambda}$)
for each group found with PM. Bottom: Parameter
estimates for the fungus network: $\lambda_{q\ell}=$ mean number
of shared host species, $\alpha_q = $ group proportion (\%) with
PM (no covariate). * stand for $\lambda_{q\ell}=$ lower than
5e--3} \label{Tab:FungusPM}
\begin{tabular*}{\textwidth}{@{\extracolsep{\fill}}ld{1.2}d{1.2}d{1.2}d{1.2}d{1.2}d{2.2}d{2.2}d{2.2}d{2.2}@{}}
\hline
& \multicolumn{1}{c}{\textbf{F1}} & \multicolumn{1}{c}{\textbf{F2}} & \multicolumn{1}{c}{\textbf{F3}} & \multicolumn{1}{c}{\textbf{F4}}
& \multicolumn{1}{c}{\textbf{F5}}
& \multicolumn{1}{c}{\textbf{F6}} & \multicolumn{1}{c}{\textbf{F7}} & \multicolumn{1}{c}{\textbf{F8}} & \multicolumn{1}{c@{}}{\textbf{F9}} \\
\hline
Size & 2 & 3 & 5 & 6 & 7 & 19 & 24 & 26 & 62 \\
$\bar{\lambda}$ & 1.68 & 1.95 & 1.65 & 0.59 & 0.85 & 0.57 & 0.50 &
0.20 & 0.12\\[6pt]
$\widehat{\lambda}_{q\ell}$  \\
F1 & 5.87 & 7.43 & 7.64 & 2.24 & 3.26 & 2.88 & 1.70 & 0.96 & 0.47 \\
F2 & 7.43 & 9.88 & 7.29 & 3.59 & 4.45 & 1.54 & 2.77 & 1.03 & 0.71 \\
F3 & 7.64 & 7.29 & 12.1 & 4.18 & 1.54 & 3.59 & 0.31 & 1.47 & 0.09 \\
F4 & 2.24 & 3.59 & 4.18 & 2.92 & 0.50 & 0.47 & 0.05 & 0.81 & 0.03 \\
F5 & 3.26 & 4.45 & 1.54 & 0.50 & 2.66 & 0.41 & 1.91 & 0.17 & 0.38 \\
F6 & 2.88 & 1.54 & 3.59 & 0.47 & 0.41 & 2.35 & * & 0.32 & * \\
F7 & 1.70 & 2.77 & 0.31 & 0.05 & 1.91 & * & 1.61 & 0.01 & 0.18 \\
F8 & 0.96 & 1.03 & 1.47 & 0.81 & 0.17 & 0.38 & 0.01 & 0.25 & *\\
F9 & 0.47 & 0.71 & 0.09 & 0.03 & 0.38 & * & 0.18 & * & 0.13\\[6pt]
$\widehat{\alpha}_q$ & 1.3 & 2.0 & 3.3 & 3.9 & 4.6 & 12 & 16 & 17 &
40\\
\hline
\end{tabular*}\vspace*{-14pt}
\end{table}
\begin{table}[h]
\caption{Fungus interaction network. Effect of different factors on the similarity of host ranges
between fungal species. $\Delta$ICL is the gain (in
log-likelihood units) obtained when switching from the best PM model to
the best PRMH model for a given covariate} \label{Tab:Conc2}
\begin{tabular}{@{}lcccc@{}}
\hline
{\textbf{Factor}} & \textbf{Covariate} & \textbf{Nb. of groups (PM)} &
\textbf{Nb. of groups (PRMH)} & $\bolds\Delta\bolds{\mathit{ICL}}$\\
\hline
Phylogenetic & Taxonomic distance & 9 & 9 & NA \\
\quad relatedness & & & & \\
[3pt]
Nutritional & Trivial distance & 9 & 9 & NA \\
\quad strategy & & & &\\
\hline
\end{tabular}
\end{table}
\end{appendix}

\section*{Acknowledgments}
We thank the D\'epartement Sant\'e des For\^ets (DSF) of the French
Minist\`ere de l'Agriculture et de la P\^eche for allowing us to use
their database. We thank Dominique Piou and Marie-Laure
Desprez-Loustau for checking the data and for helpful comments on the
results.

\begin{supplement}[id=suppA]
\stitle{Interaction network between tree and fungal species\\}
\slink[doi]{10.1214/07-AOAS361SUPP}
\slink[url]{http://lib.stat.cmu.edu/aoas/361/supplement.csv}
\sdatatype{.csv}
\sdescription{
This file contains:
\begin{itemize}
\item The adjacency matrix of interactions between tree and fungal
species.
\item The list of the tree species.
\item The list of the fungal species.
\item The matrix of genetic distances between tree species.
\item The matrix of geographical distances between tree species.
\item The matrix of taxonomic distances between fungal species.
\item The matrix of nutritional type of the fungal species.
\end{itemize}
}
\end{supplement}


\printaddresses

\end{document}